\documentclass{ifacconf}

\usepackage{amsmath,amsfonts,amssymb}
\usepackage{graphicx}      
\usepackage{graphics}
\usepackage{natbib}        
\usepackage{color}

\newcommand{\be}{\begin{equation}}
\newcommand{\ee}{\end{equation}}
\newcommand{\bd}{\begin{displaymath}}
\newcommand{\ed}{\end{displaymath}}
\newcommand{\bea}{\begin{eqnarray}}
\newcommand{\eea}{\end{eqnarray}}

\newcommand{\R}{\mathbb{R}}
\newcommand{\C}{\mathbb{C}}
\newcommand{\Z}{\mathbb{Z}}

\newcommand{\Cs}{\mathbb{C}_{\sigma_0}}

\newcommand{\w}{\omega}
\newcommand{\so}{\sigma_0}
\newcommand{\dwz}{\Delta\w_{zr}}
\newcommand{\dwp}{\Delta\w_{pi}}
\newcommand{\dsz}{\Delta \sigma_{zr}}
\newcommand{\dsp}{\Delta \sigma_{pi}}

\makeatletter
\def\Ddots{\mathinner{\mkern1mu\raise\p@
\vbox{\kern7\p@\hbox{.}}\mkern2mu
\raise4\p@\hbox{.}\mkern2mu\raise7\p@\hbox{.}\mkern1mu}}
\makeatother

\begin{document}
\begin{frontmatter}

\title{Continuation Based Computation of Root-Locus for SISO Dead-Time Systems\thanksref{sponsors}}

\author[First]{Suat Gumussoy},
\author[First]{Wim Michiels}

\address[First]{Department of Computer Science, K. U. Leuven, \\
        Celestijnenlaan 200A, 3001, Heverlee, Belgium \\
        \mbox{(e-mails: \{suat.gumussoy, wim.michiels\}@cs.kuleuven.be)}.}

\thanks[sponsors]{This article present results of the Belgian Programme on Interuniversity Poles of Attraction, initiated by the
Belgian State, Prime Minister's Office for Science, Technology and Culture, the Optimization in
Engineering Centre OPTEC of the K.U.Leuven, and the project STRT1-09/33 of the K.U.Leuven Research Foundation.}

\begin{abstract}
We present a numerical method to plot the root-locus of Single-Input-Single-Output (SISO) dead-time systems on a given right half-plane up to a predefined controller gain. We compute the starting and intersection points of root-locus inside the region and we obtain the root-loci of each root based on a predictor-corrector type continuation method. The method is effective for high-order SISO dead-time systems.
\end{abstract}


\end{frontmatter}

\section{Introduction}
The root-locus method is an essential tool in modern control engineering for analysis and synthesis problems, \cite{OgataBook}. This method is successfully implemented for finite dimensional SISO systems and becomes a fundamental tool in control education, \cite{Evans04, Krajewski07}.

The closed-loop of the SISO system with a time-delay and a constant gain has infinitely many poles in the complex-plane, \cite{WimBook}. Therefore the root-locus plot for dead-time systems is a difficult problem. Unlike the finite dimensional case, the root-locus equation contains a time-delay term and standard polynomial root-finding algorithms for the root computation are not available.

There are several approaches to plot the root-locus of SISO dead-time systems. The approximate root-locus can be plotted using the classical root-locus technique and the Pad\'{e} approximation of time-delay term, \cite{Ozbay06}. This approach guarantees the accuracy of the approximation only inside a certain region and has numerical problems when high order Pad\'{e} approximation is needed for larger regions. The direct approaches to which our method belongs, obtain the root-locus plot without approximating the time-delay term. A graphical method based on the computation of the root-locus gain and the closed-loop roots on various vertical lines in the complex-plane is presented in \cite{HuangTAC67}. The root-locus on a rectangular region in the complex-plane is obtained by finding points on a rectangular grid satisfying the root-locus equation and connecting these points, \cite{Krall70}. A continuation based method computes the root-locus on a rectangular region on the complex plane using the slope of the phase equation of the root-locus equation to determine a prediction step direction and a Newton-Raphson iteration to correct the prediction values, \cite{AshTAC68}. The method detects the roots entering into the region by checking the sign of function values on the constant grid points of the region's boundary.

The root-locus equation of a SISO dead-time system is written as two real-valued equations, \cite{Yeung82}. For a given constant imaginary part, the real part of the root on the root-locus is computed by finding the roots of a polynomial. By sweeping various values on the imaginary axis, the root-locus plot is obtained.

In \cite{Suh82}, a continuation method for the root-locus analysis, \cite{PanChao78}, is extended to SISO dead-time systems. The root-loci branch is followed by computing the solution of non-linear differential equations with respect to the controller gain. The root-locus equation is transformed into another root-locus equation whose starting points are computed easily. Using differential equations for the second root-locus equation, these points are followed until the roots of both root-locus equations coincide. Since the common points are the initial points of the original root-locus equation, the root-locus plot is obtained for the original root-locus equations using its corresponding differential equations. In \cite{Nishioka91} the same approach is used to compute the initial points of the roots entering into the region. Instead of computing the root-locus by solving differential equations, the controller gain is written as a function of other terms in the root-locus equation and the zeros of the imaginary part of the controller gain is computed by a triangulation method on the complex plane. The last two approaches are applicable to SISO time-delay systems with state-delays. However these methods follow the root-locus trajectories with respect to the controller gain which is numerically ill-posed due to the high sensitivity in the neighborhood of intersection points, which are characterized by the presence of multiple roots. The detection of asymptotic roots requires solving another root-locus problem and the number of required roots in a complex region is difficult to estimate.

In this paper, we compute the root-locus plot of SISO dead-time systems on a given complex right half-plane up to a predefined controller gain. We calculate the starting points, the poles of SISO dead-time system and the roots entering into the region, and branching points of the root-locus inside the region. We follow the root trajectory based on a predictor-corrector type continuation method.

Our main contributions are the following:
\begin{itemize}
\item We compute all the roots entering into the region and their controller gains for an upper bound controller gain. By choosing the controller gain sufficiently large, the asymptotic behavior of the roots can be seen.
\item Our continuation method estimates the next root on the complex plane by a simple linear predictor and corrects this prediction with a Newton method. Since the trajectory following method is based on a parameterization of curves in the (root,gain) space in terms of arclength, it is numerically robust.
\item We use an adaptive step size in the prediction step depending on the convergence rate of the Newton method and the distance of the root from the root-locus trajectory. This makes our algorithm scalable by using different step sizes for different root-locus trajectories.
\item Most of the methods in the literature requires many evaluations of the transfer function of the SISO dead-time system. The evaluation of the finite-dimensional transfer function is numerically expensive and the function values are not numerically stable due to the oscillation and exponential increase of the time-delay term in the imaginary axis and the positive real axis direction in the complex-plane respectively. We avoid these problems by transforming the root-locus equation into phase and logarithmic magnitude equations and evaluating in a numerically stable way. We need these evaluations only in the correction step.
\end{itemize}

The paper is organized as follows. In Section \ref{sec:prob} we formulate the problem. The starting and branching points of root-locus inside the region are computed in Section \ref{sec:critpts}. The predictor-corrector based continuation method is given in Section \ref{sec:rloci}. The overall algorithm for root-locus plot is presented in Section \ref{sec:alg}. Section~\ref{sec:ex} is devoted to a numerical example. In Section~\ref{sec:concl} some
concluding remarks are presented.

\vspace{-.1cm}
\textbf{Notation:} \\
\begin{tabbing}
  \= $\C, \R, \Z$ \=: fields of complex, real and integer numbers, \\
  \> $\Re(u)$ \>: real part of a complex number $u$, \\
  \> $\Im(u)$ \>: imaginary part of a complex number $u$, \\
  \> $|u|, \angle u$ \>: magnitude and phase of a complex number $u$, \\
  \> $\lfloor u \rfloor$ \>: the next smallest integer close to a real number $u$, \\
  \> $\lceil u \rceil$ \>: the next largest integer close to a real number $u$, \\
  \> $u^T$ \>: the transpose of the vector $u$, \\
  \> $\textrm{sign}(u)$ \>: returns $+1, -1, 0$ given a real number $u$ \\
  \>  \>$\ $ for $u>0, u<0, u=0$ respectively. \\
\end{tabbing}
\section{Problem Formulation} \label{sec:prob}
A SISO dead-time system is a proper SISO system with a constant input or output time-delay, $h\in\R, h>0$ and it has transfer function representation
\be \label{eq:tfG}
{\textstyle G(s)e^{-hs}=\alpha\frac{\prod_{r=1}^m s-(\sigma_{zr}+j\omega_{zr})}
{\prod_{i=1}^n s-(\sigma_{pi}+j\omega_{pi})}e^{-hs}}
\ee where $\alpha\in\R$, $\sigma_{zr}+j\omega_{zr}$ $r=1,\ldots,m$, $\sigma_{pi}+j\omega_{pi}$ $i=1,\ldots,n$ are the system gain, zeros and poles of $G$ respectively. The \emph{root-locus equation} of a SISO dead-time system is
\be \label{eq:rlocus}
{\textstyle f(s,k)=1+kG(s)e^{-hs}=0}
\ee where $k\in\R$, $k\geq0$ is the controller gain. We define the \emph{root-locus region} as
\be \label{eq:Cs}
{\textstyle \Cs =\left\{s\in \mathbb{C}: \Re(s)\geq\sigma_0\right\}}
\ee and the \emph{boundary} of the root-locus region is a \emph{given} vertical line parallel to the imaginary axis, $\Re(s)=\sigma_0<0$,  $\sigma_0\in\R$ and its value depends on the analysis requirements. We consider the following root-locus problem: \\
\noindent \textbf{Problem:} Compute the root-locus of the SISO dead-time system (\ref{eq:tfG}) inside the root-locus region $\Cs$ for $k\in[0,k_{\max}]$ where $k_{\max}$ is a given positive real number.

When a root $s$ of (\ref{eq:rlocus}) crosses the boundary $\Re(s)=\so$ at the controller gain $k$, we determine whether it enters into or leaves the region $\Cs$ by computing its \emph{crossing direction}, \cite{WimBook} defined as
\be \label{eq:RT}
{\textstyle
\mathcal{CD}(s,k):=\textrm{sign}\left(\Re{\left(\left.-\frac{\frac{\partial f}{\partial k}}{\frac{\partial f}{\partial s}}\right|_{f(s,k)=0}\right)}\right).
}\ee
Note that a root on the boundary $\Re(s)=\so$ enters into or leaves the region $\Cs$ when $\mathcal{CD}(s,k)>0$ or $\mathcal{CD}(s,k)<0$ respectively.
\section{Computation of Critical Points of Root-Locus} \label{sec:critpts}
The \emph{critical points} of root-locus are:
\begin{itemize}
\item the starting points of the root-locus, the poles of $G$ inside $\Cs$, and the roots of (\ref{eq:rlocus}) crossing the boundary of the root-locus region $\Re(s)=\so$ for some $k\in[0,k_{\max}]$,
\item the branching points of the root-locus where two or more root-locus trajectories intersect inside the region~$\Cs$.
\end{itemize}
The computation of roots of (\ref{eq:rlocus}) crossing the boundary $\Re(s)=\so$ and their crossing directions are given in Section~\ref{sec:bndx}.

A branching point $s$ satisfies the root-locus equation (\ref{eq:rlocus}) and
\be
{\textstyle \frac{\partial f(s,k)}{\partial s}=k(G'(s)-G(s)h)e^{-hs}=0.}
\ee Thus, the branching points are the zeros of $G'(s)-G(s)h$ inside the region $\Cs$, satisfying the root-locus equation (\ref{eq:rlocus}) for a controller gain $k\in\R$, $k>0$. Since $G'(s)-G(s)h$ is a rational transfer function, its zeros can be computed by standard polynomial root-finding algorithms and the branching points can be determined accurately. The behavior of a root around a branching point is given in the following Lemma, \cite{Suh82}.
\begin{lem} \label{lem:branch}
Assume that $\tilde{s}$ is a root of the root-locus equation for a controller gain $\tilde{k}$, i.e., $f(\tilde{s},\tilde{k})=0$ with multiplicity $N$, i.e.,
\bd
{\textstyle
\left.\frac{\partial^l f(s,\tilde{k})}{\partial s}\right|_{s=\tilde{s}}=0, \ l=1,\ldots,N-1,\ \textrm{and} \ \left.\frac{\partial^N f(s,\tilde{k})}{\partial s}\right|_{s=\tilde{s}}\neq0.}
\ed
\end{lem}
Then the root-locus has $N$ intersecting trajectories at $s=\tilde{s}$ and the angle of direction change of a root incoming to and going from a branching point is $0$ or $-\frac{\pi}{N}$ when $N$ is odd or even respectively.

Since branching points inside the region $\Cs$ and their multiplicities are computed before-hand, Lemma \ref{lem:branch} allows us to determine the direction of a root-locus trajectory after a branching point.
\subsection{Computation of Roots Crossing the Boundary of the Root-Locus Region} \label{sec:bndx}
A root $s$  on the boundary of root-locus region $\Re(s)=\so$ for the controller gain $k$ satisfies the magnitude and phase equations of the root-locus equation (\ref{eq:rlocus}). We first find the intervals on the boundary where the magnitude condition holds for some $k\in[0,k_{\max}]$. This is equivalent to finding the intervals on $\Re(s)=\so$ and $\w\in[0,\infty)$ such that
\be \label{eq:Kcond}
{\textstyle K(\w):=h\so-\ln|G(\so+j\w)|\leq\ln k_{\max}.}
\ee
\begin{lem} \label{lem:polyK}
Assume that $G$ has no poles or zeros on the boundary of the root-locus region. The functions $K(\w)$ and $K'(\w)$ are continuous and the non-negative zeros of $K'(\w)$ are the non-negative real roots of the polynomial,
\be \label{eq:polyKp}
\Gamma_z(\w)\sum_{i=1}^n\dwp\Gamma_p^i(\w)-\Gamma_{p}(\w)\sum_{r=1}^m\dwz\Gamma_z^r(\w)
\ee
where $\dsz=(\so-\sigma_{zr})$, $\dwz=(\w-\w_{zr})$, $\gamma_{zr}(\w)=\dsz^2+\dwz^2$, $\Gamma^r_z(\w)=\prod_{\substack{r_1=1\\r_1\neq r}}^m\gamma_{zr}(\w)$ for $r=1,\ldots,m$, $\dsp=(\so-\sigma_{pi})$, $\dwp=(\w-\w_{pi})$, $\gamma_{pi}(\w)=\dsp^2+\dwp^2$, $\Gamma^i_p(\w)=\prod_{\substack{i_1=1\\i_1\neq k}}^n\gamma_{pi}(\w)$ for $i=1,\ldots,n$, $\Gamma_z(\w)=\prod_{r=1}^m\gamma_{zr}(\w)$, $\Gamma_p(\w)=\prod_{i=1}^n\gamma_{pi}(\w)$.
\end{lem}

\noindent\textbf{Proof.\ }
Using the transfer function of G (\ref{eq:tfG}), the function $K(\w)$ can be written as,
\be \label{eq:Kw}
{\textstyle K(\omega)=h\so-\ln|\alpha|+\frac{1}{2}\left(\sum_{i=1}^n
\ln\gamma_{pi}(\omega)-\sum_{r=1}^m
\ln\gamma_{zr}(\omega)\right).}
\ee
The first derivative of $K(\w)$ (\ref{eq:Kw}) is
\be
{\textstyle K'(\w)=\sum_{i=1}^n \frac{\dwp}{\gamma_{pi}(\omega)}-\sum_{r=1}^m \frac{\dwz}{\gamma_{zr}(\omega)} \label{eq:Kwp}.}
\ee

The functions $K(\w)$ and $K'(\w)$ are continuous except the points where $\gamma_{zr}(\w)$ or $\gamma_{pi}(\w)$ are equal to zero. These points are the poles or zeros of $G$ on $\Re(s)=\so$. The continuity results in Lemma \ref{lem:polyK} follow from the assumption. The polynomial (\ref{eq:polyKp}) is the numerator of the function $K'(\w)$ in (\ref{eq:Kwp}) and the result follows.  \hfill $\Box$

\begin{cor} \label{cor:Kmonotonic}
Assume that $G$ has no poles or zeros on the boundary of the root-locus region. Then function $K(\w)$ is monotonic on the intervals whose boundary points (without multiplicity) are successive non-negative zeros of $K'(\w)$, $0$ and $\infty$.
\end{cor}
\noindent\textbf{Proof.\ }
By Lemma \ref{lem:polyK}, the function $K(\w)$ is continuous since $\so$ is chosen such that there are no poles or zeros of $G$ on $\Re(s)=\so$. Therefore it is monotonic inside the intervals determined by its extremum points and the end points of the boundary of the root-locus region, $0$ and $\infty$.~$\Box$

Since $K(\w)$ is monotonic on each interval in Corollary \ref{cor:Kmonotonic}, we find the subinterval in each interval where $K(\w)$ satisfies (\ref{eq:Kcond}). This is done as follows. If the values of $K(\w)$ at the interval end points are smaller than $\ln k_{\max}$, then all $K(\w)$ values in this interval are smaller than $\ln k_{\max}$ because $K(\w)$ is monotonic. If one of the values of $K(\w)$ at the interval end points is larger and the other one is smaller than $\ln k_{\max}$, we can find the point where $K(\w)$ is equal to $\ln k_{\max}$ by a bisection algorithm and take the subinterval satisfying (\ref{eq:Kcond}). If both values of $K(\w)$ at the interval end points are larger than $\ln k_{\max}$, we discard that interval since all values of $K(\w)$ are larger than $\ln k_{\max}$ and the condition (\ref{eq:Kcond}) never holds. Based on this approach, we can compute the set of intervals $I$ on the boundary of the root-locus region where the magnitude condition (\ref{eq:Kcond}) is satisfied for some values of $0\leq k\leq k_{\max}$.

The roots crossing the boundary of the root-locus region also satisfy the phase equation of (\ref{eq:rlocus}):
\be \label{eq:Phcond}
(2l+1)\pi=\phi(\w),\ \   l\in\Z,
\ee
over the intervals $I$ on $\Re(s)=\so$. Here the function $\phi(\w)$ represents the \emph{continuous} extension of the phase of the transfer function $G(s)e^{-hs}$ and their equivalence is
\bd
{\textstyle \mod\left(\angle \left.G(s)e^{-hs} \right|_{s=\so+j\w},2\pi\right)=\mod\left(\phi(\w),2\pi\right)}
\ed
where $\w\in[0,\infty)$. The left hand-side of the equation (\ref{eq:Phcond}) represents constant functions of $\omega$. If we partition the intervals $I$ into the subintervals such that the function $\phi(\w)$ is monotonic on each subinterval, we can compute the boundary crossing roots by a bisection algorithm. The following results allow us to compute the intervals on $\Re(s)=\so$ where the function $\phi(\w)$ is monotonic.

\begin{lem} \label{lem:polyPhi}
Assume that $G$ has no poles or zeros on the boundary of the root-locus region. Then the functions $\phi(\w)$ and $\phi'(\w)$ are continuous and the non-negative zeros of $\phi'(\w)$ are the non-negative real roots of the polynomial,
\be
 \label{eq:polyPhi}
 \Gamma_{p}(\w)\sum_{r=1}^m\dsz\Gamma_z^r(\w)-\Gamma_z(\w)\sum_{i=1}^n\dsp\Gamma_p^i(\w)-h\Gamma_z(\w)\Gamma_{p}(\w) \ee
where the functions $\Gamma^r_z(\w)$ for $r=1,\ldots,m$, $\Gamma^i_p(\w)$ for $i=1,\ldots,n$, $\Gamma_z(\w)$ and $\Gamma_p(\w)$ are defined in Lemma \ref{lem:polyK}.
\end{lem}
\noindent\textbf{Proof.\ }
Using the transfer function of $G$ (\ref{eq:tfG}), the function $\phi(\w)$ is written as
\be \label{eq:Phi}
\phi(\w)=\phi_1(\w)+\phi_0
\ee where
\be \label{eq:Phi1}
{\textstyle  \phi_1(\w):=\sum_{r=1}^m \tan^{-1} \frac{\dwz}{\dsz}
-\sum_{i=1}^n \tan^{-1}\frac{\dwp}{\dsp}
-h\w}
\ee and $\phi_0$ is the offset difference, $0$ or $\pi$ between $\phi(\w)$ and $\phi_1(\w)$ (\ref{eq:Phi1}) defined as $\phi_0=\angle{G(\sigma_0)}-\phi_1(0)$.

The first derivative of the function $\phi(\w)$ is
\be \label{eq:Phip}
{\textstyle  \phi'(\w)=\sum_{r=1}^m \frac{\dsz}{\gamma_{zr}(\w)}-
\sum_{i=1}^n \frac{\dsp}{\gamma_{pi}(\w)}-h.}
\ee

Following the same arguments in Lemma \ref{lem:polyK}, the functions $\phi(\w)$ and $\phi'(\w)$ are continuous by the assumption. The polynomial (\ref{eq:polyPhi}) is the numerator of the function $\phi'(\w)$ (\ref{eq:Phip}) and the result follows. \hfill $\Box$

\begin{cor} \label{cor:polyPhimonotonic}
Assume that $G$ has no poles or zeros on the boundary of the root-locus region. Then the function $\phi(\w)$ is monotonic on each interval in the set of intervals $I_\phi$  whose boundary points (without multiplicity) are successive non-negative zeros of $\phi'(\w)$, $0$ and $\infty$.
\end{cor}
\noindent\textbf{Proof.\ }
The function $\phi(\w)$ is continuous. The monotonicity of $\phi(\w)$ changes only at the points where $\phi'(\w)=0$. The assertion follows. \hfill $\Box$

The intersection of two sets of intervals $I$ and $I_\phi$ partitions $I$ into the subintervals, $I=\cup_{i=1}^{n_I}I_i$, where $\phi(\w)$ is monotonic on each interval $I_i$. Each intersection of the function $\phi(\w)$ and the constant functions in the left hand side of (\ref{eq:Phcond}) over the intervals $I$ corresponds to a boundary crossing root since any such point on $I$ satisfies both the magnitude
condition (\ref{eq:Kcond}) and the phase equation of the root-locus equation (\ref{eq:Phcond}) on $\Re(s)=\so$. Since the function $\phi(\w)$ is monotonic on $I_i$, we can compute each intersection point by a bisection algorithm for $\phi(\w)$ over the interval~$I_i$. The value of the $\w$ at the intersection point is the imaginary part of the boundary crossing root on the interval $I_i$ and the corresponding controller gain is the value of $K(\w)$ for this point. If there is no horizontal line intersecting $\phi(\w)$ on $I_i$, we discard the interval since there is no root crossing this interval. We compute the roots of (\ref{eq:rlocus}) crossing the boundary of the root-locus region by the following algorithm.

\begin{alg} \label{eq:bndxroot}
${}$\\
For each interval in $I_i=[\w_i^L,\w_i^R]$ of $I=\cup_{i=1}^{n_I}I_i$,
\begin{enumerate}
\item Compute $\phi_i^{\max}$, $\phi_i^{\min}$, the maximum and minimum of $\phi(\w_i^L)$, $\phi(\w_i^R)$.
\item Compute $l_i^{\max}=\left\lfloor \frac{\phi_i^{\max}}{2\pi}-\frac{1}{2} \right\rfloor$ and $l_i^{\min}=\left\lceil \frac{\phi_i^{\min}}{2\pi}-\frac{1}{2} \right\rceil$.
\item If $(l_i^{\min}>l_i^{\max})$ discard the interval $I_i$, \\
else \\
for $l=l_i^{\min}$ to $l_i^{\max}$
\begin{itemize}
\item find the imaginary part of the boundary crossing root $\w_{cr}$ at the intersection of the horizontal line $(2l+1)\pi$ and $\phi(\w)$ (\ref{eq:Phi}) by a bisection algorithm  over the interval $I_i$.
\item compute the corresponding controller gain for the boundary crossing root, $K_{cr}=K(\w_{cr})$.
\end{itemize}
\end{enumerate}
\end{alg}

Note that the controller gain $K_{cr}$ is equal to $K_{cr}=\ln k_{cr}$ where $k_{cr}$ is controller gain in the root-locus equation (\ref{eq:rlocus}). In the remainder of the paper, we use capital $K$ and the small $k$ for the controller gain in logarithmic base and the original one in (\ref{eq:rlocus}).

By Algorithm~\ref{eq:bndxroot}, we compute all roots crossing the boundary of the root-locus region $\Re(s)=\so$ and the corresponding controller gain values for $k\in[0,k_{\max}]$. The crossing directions of these roots are determined using the following theorem.

\begin{thm} \label{thm:CD}
The crossing direction of a boundary crossing root, $s_{cr}=\so+j\w_{cr}$ only depends on the imaginary part $\w_{cr}$ on the boundary of root-locus region $\Re(s)=\so$ and is equal to \vspace{-.4cm}
\be
{\textstyle  \mathcal{CD}(s_{cr},k_{cr})=-\textrm{sign}\left(\phi'(\w_{cr})\right).}
\ee
\end{thm}

\noindent\textbf{Proof.\ }
Using the transfer function representation in (\ref{eq:tfG}) and (\ref{eq:Kwp},\ref{eq:Phip}), we obtain
\be \label{eq:GpG}
{\textstyle  G'(s_{cr})G^{-1}(s_{cr})-h=\phi'(\w_{cr})+j K'(\w_{cr}).}
\ee
By (\ref{eq:RT}) and (\ref{eq:GpG}), the crossing direction of $s_{cr}$ at $k=k_{cr}$ is equal to
\bd
{\textstyle  \mathcal{CD}(s_{cr},k_{cr})=\textrm{sign}\left(\Re\left(\left(k_{cr}
\left(h-\frac{G'(s_{cr})}{G(s_{cr})}\right)\right)^{-1}\right)\right)}
\ed
\bd
\hspace{2.5cm} {\textstyle =-\textrm{sign}(\phi'(\w_{cr})).} \hspace*{3.4cm} \Box
\ed
By Theorem \ref{thm:CD}, the crossing directions of roots crossing $\Re(s)=\so$ are the same when their imaginary parts are inside the same interval of $I_\phi$ in Corollary \ref{cor:polyPhimonotonic}. Using this result, we determine the crossing directions of boundary crossing roots from their imaginary parts. We group the boundary crossing roots according to their crossing directions and define the sets $W^{in}$ and $W^{out}$ as
\bd
{\textstyle W^{in}=\{s_\nu^I ,K_\nu^I\}_{\nu=1}^{n_i}\ \textrm{and}\ W^{out}=\{s_\nu^O,K_\nu^O\}_{\nu=1}^{n_o}}
\ed where $s_\nu^I=\so+j\w_\nu^I, K_\nu^I$ for $\nu=1,\ldots,n_i$, $s_\nu^O=\so+j\w_\nu^O, K_\nu^O$  for $\nu=1,\ldots,n_o$ are the boundary crossing roots entering into or leaving the root-locus region and their controller gains respectively.

\noindent\textbf{Remark 1:\ } The crossing direction formula (\ref{eq:RT}) is well-posed (either $+1$ or $-1$) if there are no poles or zeros of $G$ or branch points on the boundary of the root-locus region.\\
\noindent\textbf{Remark 2:\ } If the plant $G$ is bi-proper (i.e., $d:=G(\infty)\neq0$), then the controller gain must be bounded as $k_{\max}<\frac{e^{h\so}}{|d|}$. For larger controller gains, the root-locus region $\Cs$ always has infinitely many roots.

\section{Computing a Root-Locus Trajectory} \label{sec:rloci}

The starting points of the root-locus are the poles of $G$ inside $\Cs$ at $k=0$ and the roots of the root-locus equation~(\ref{eq:rlocus}) entering into the root-locus region, $\so+j\w_\nu^I$ at $k=k_\nu^I:=e^{K_\nu^I}$ for $\nu=1,\ldots,n_i$. We compute each root-locus trajectory by a secant-predictor, Newton-corrector continuation method, \cite{ContinuationSIAMBook}. In the prediction step, a line passing through the last two computed roots and controller gains is used to estimate the next root and controller gain at a certain distance (steplength) in the (root,gain) parameter space. This estimate is corrected using Newton's method in the correction step. The next iteration continues in a similar way, though the step length is adaptive.

\subsection{Prediction Step}
The predicted root and the controller gain computation in the prediction step requires the previous root, the controller gain, a direction and a step length. For each root-locus trajectory, the starting point $s_0$ is available. The direction of the prediction step $d_i$ is computed as follows:
\begin{itemize}
\item Initial directions $d_0^s\in\C$ for the poles of $G$ inside $\Cs$ and the boundary crossing roots are computed by the phase equation of root-locus (\ref{eq:rlocus}) and by the phase of the derivative of the roots with respect to the controller gain for boundary crossing roots, i.e.,
\bea
\nonumber {\textstyle\left.\frac{\partial s}{\partial k}\right|_{(s,k)=(s_l^i, k_l^i)}}&=&{\textstyle -\left.\left(k\left(\frac{G'(s)}{G(s)}-h\right)\right)^{-1} \right|_{(s,k)=(s_l^i, k_l^i)},} \\
\nonumber &=&{\textstyle-\left(k_l^i\left(\phi(\w_l^i)+j K(\w_l^i)\right)\right)^{-1}.}
\eea Set the root-locus direction as $d_i=\left[
                                            \begin{array}{ccc}
                                              \Re(d_0^s) & \Im(d_0^s) & 1 \\
                                            \end{array}
                                          \right]^T$ and normalize to $1$.
\item The directions in other iterations are computed using the real and imaginary parts of the last two corrected roots and the controller gains,
$\tilde{s}_{i}^c=
\left[
\begin{array}{ccc}
\sigma_i^c & \w_i^c & K_i^c
\end{array}\right]^T$,
$\tilde{s}_{i-1}^c=
\left[
\begin{array}{ccc}
\sigma_{i-1}^c & \w_{i-1}^c & K_{i-1}^c
\end{array}\right]^T$, as
\be \label{eq:preddirection}
{\textstyle d_i=\frac{\tilde{s}_i^c-\tilde{s}_{i-1}^c}{\|\tilde{s}_i^c-\tilde{s}_{i-1}^c\|},\ i\geq 1.}
\ee
\end{itemize}
The real and imaginary parts of the predicted root and the controller gain $\tilde{s}_{i+1}^p=\left[
\begin{array}{ccc}
\sigma_{i+1}^p & \w_{i+1}^p & K_{i+1}^p)
\end{array}\right]^T$
are computed using a line equation with a step length $h_i$
\be \label{eq:secpred}
\tilde{s}_{i+1}^p=\tilde{s}_i^c+d_i h_i,\ i\geq 0.
\ee The initial step length $h_0$ is fixed. The step lengths in other iterations are calculated adaptively based on previous values, as we outline later on.
\subsection{Correction Step}
We use Newton's method to solve a set of nonlinear equations to find the real and imaginary parts of the corrected root and the corrected controller gain $\tilde{s}_{i+1}^c=\left[
\begin{array}{ccc}
\sigma_{i+1}^c & \w_{i+1}^c & K_{i+1}^c)
\end{array}\right]^T$.  These equations are given by
\bea
{\textstyle M(\sigma_{i+1}^c,\w_{i+1}^c,K_{i+1}^c)}&=&{\textstyle 0} \label{eq:MNew} \\
{\textstyle P(\sigma_{i+1}^c,\w_{i+1}^c)}&=&{\textstyle 0} \\
{\textstyle (\tilde{s}_{i+1}^c-\tilde{s}_{i+1}^p) d_i}&=&{\textstyle 0} \label{eq:line}
\eea where
\bea
\nonumber {\textstyle M(\sigma,\w,K)}&=&{\textstyle \ln|\alpha|+\frac{1}{2}\sum_{r=1}^m \left(\ln (\sigma-\sigma_{zr})^2+(\w-\w_{zr})^2\right)} \\
 &&\hspace{-1,5cm} {\textstyle -\frac{1}{2}\sum_{i=1}^n \left(\ln (\sigma-\sigma_{pi})^2+(\w-\w_{pi})^2\right)-h\sigma+K,} \label{eq:M}\\
\nonumber {\textstyle P(\sigma,\w)}&=&{\textstyle \angle \alpha+\sum_{r=1}^m \tan^{-1}\frac{\w-\w_{zr}}{\sigma-\sigma_{zr}}-\sum_{i=1}^n \tan^{-1}\frac{\w-\w_{pi}}{\sigma-\sigma_{pi}}} \\
&& \hspace{2cm}   {\textstyle -h\w-\pi}, \label{eq:P}
\eea and $P(\sigma,\w)$ has a range $(-\pi,\pi]$.

The functions $M$ (\ref{eq:M}) and $P$ (\ref{eq:P}) are equivalent representations of the magnitude and phase equations of the root-locus equation (\ref{eq:rlocus}). The arctangent functions in $P(\sigma,\w)$ are implemented as two argument function atan2 with the range $(-\pi,\pi]$. The equation (\ref{eq:line}) guarantees that the (linearized) distance of the corrected root and the controller gain $\tilde{s}_{i+1}^c$ from the predicted root and the controller gain $\tilde{s}_{i+1}^p$ is equal to the step size $h_i$.

Based on the set of equations in (\ref{eq:MNew}-\ref{eq:line}), we implement Newton's method as
\be
{\textstyle (J_{i+1}^m)\left(\tilde{s}_{i+1}^{m+1}-\tilde{s}_{i+1}^m\right)=-\tilde{f}_{i+1}^m,} \ m=0,1,\ldots,m_{\tilde{s}}
\ee where $\tilde{s}_{i+1}^m=\left[\begin{array}{ccc} \sigma_{i+1}^{m} & \w_{i+1}^{m} & K_{i+1}^m \end{array}\right]^T$ is the vector of the real and imaginary part of the corrected root and the controller gain at $m^\textrm{th}$ Newton iteration. The function $\tilde{f}_{i+1}^m$ and its Jacobian $J_{i+1}^m$ are defined as
\bd
{\textstyle \tilde{f}_{i+1}^m=
\left(
\begin{array}{l}
M(\sigma,\w,K) \\
P(\sigma,\w) \\
(\tilde{s}_{i+1}^m-\tilde{s}_{i+1}^p) d_i-h_i
\end{array}\right), {\small J_{i+1}^m = \left(\begin{array}{ccc}
\frac{\partial M}{\partial \sigma} & \frac{\partial M}{\partial \w} & \frac{\partial M}{\partial K} \\
& & \\
\frac{\partial P}{\partial \sigma} & \frac{\partial P}{\partial \w} & 0\\
& & \\
& d_i^T &
\end{array} \right)}}
\ed where $(\sigma,\w,K)=(\sigma_{i+1}^m,\w_{i+1}^m,K_{i+1}^m)$.
The initial point $\tilde{s}_{i+1}^0$ for the correction step is the vector of the real and imaginary part of the predicted root and the controller gain from the prediction step $\tilde{s}_{i+1}^0=\tilde{s}_{i+1}^p$.

The Newton iterations continue until the root and the controller gain converge to a point within a predefined tolerance and the corrected root and the controller gain are set to the last iteration value in Newton method, $s_{i+1}^c=\tilde{s}_{i+1}^{m_{\tilde{s}}}$.
\subsection{Adaptive Step Length}
The step length computation for the next prediction step depends on two factors, \cite{ContinuationSIAMBook}
\begin{itemize}
\item the \emph{contraction rate} of the first two successive Newton steps in the corrector step, i.e.,
\bd
{\textstyle \kappa_{i+1}:=\frac{\|J_{i+1}^0\tilde{f}_{i+1}^{1}\|}{\|J_{i+1}^0\tilde{f}_{i+1}^{0}\|};}
\ed
\item the \emph{distance} to the root-locus
\bd
{\textstyle \delta_{i+1}=\left\|1-e^{M(\sigma_{i+1}^c,\w_{i+1}^c,K_{i+1})+jP(\sigma_{i+1}^c,\w_{i+1}^c)}\right\|.}
\ed
\end{itemize}
The individual deceleration factors are calculated as
\bd
{\textstyle \kappa_{df}=\sqrt{\frac{\kappa_{i+1}}{\tilde{\kappa}}},\ \textrm{and} \
\delta_{df}=\sqrt{\frac{\delta_{i+1}}{\tilde{\delta}}}}
\ed where $\tilde{\kappa}$, $\tilde{\delta}$ are the nominal contraction rate and the distance. The overall deceleration factor of the step length is computed as
\bd
{\textstyle h_{df}:=\max\{\kappa_{df},\delta_{df}\}}
\ed and limited to $[\frac{1}{2},2]$,
\bd
{\textstyle \bar{h}_{df}:=\max\{\min\{h_{df},2\},\frac{1}{2}\}.}
\ed Note that if $\bar{h}_{df}=2$, the predictor step is repeated with a reduced step length. This check is done inside the corrector step to avoid unnecessary Newton iterations (see \cite{ContinuationSIAMBook} for further details).
The step length for the next prediction step is
\be
{\textstyle h_{i+1}={h_i}/{\bar{h}_{df}}.}
\ee

\section{Algorithm} \label{sec:alg} \label{sec:alg}
We compute the root-locus of the SISO dead-time system~(\ref{eq:tfG}) inside the root-locus region $\Cs$ by the following algorithm.

\begin{enumerate}
\item Compute the critical points of root-locus as explained in Section \ref{sec:critpts}, i.e. ~ the starting and branching points of root-locus trajectories,
\item For each root-locus trajectory computation:
\begin{enumerate}
\item Using the starting point as an initial root, compute the next root as explained in Section \ref{sec:rloci} by computing
    \begin{enumerate}
    \item the predicted root and the controller gain, $\tilde{s}_{i+1}^p$,
    \item the corrected root and the controller gain, $\tilde{s}_{i+1}^c$,
    \item the step length computation for the next prediction step, $h_{i+1}$.
    \end{enumerate}
    until that the trajectory reaches a branching point or the controller gain of the root exceeds $\ln k_{\max}$ or the trajectory leaves the root-locus region $\Cs$.
\item When the trajectory reaches a branching point, go to Step $2-a)$ and continue to compute the roots and the controller gains where the starting point is the branching point and the initial direction is calculated by Lemma \ref{lem:branch}.
\item When the controller gain exceeds $\ln k_{\max}$ or the trajectory leaves the root-locus region, stop the computation of the roots and the controller gains for this trajectory (If the trajectory leaves $\Cs$, check that it crosses one of the points in the set $W^{out}$). Go to Step $2-a)$ and start to compute another root-locus trajectory by choosing another starting point.
\end{enumerate}
\end{enumerate}

Note that if the root-locus trajectory is on the real axis, we can continue from the next branching point with the controller gain less than $\ln k_{\max}$ or compute the root on the real axis whose controller gain is equal to $\ln k_{\max}$ by a bisection algorithm.

\noindent\textbf{Remark 1:\ } The algorithm can be modified to include negative controller gains, $k\in[-k_{\max},k_{\max}]$. Then the constant functions on the left-hand side of the phase equation in (\ref{eq:Phcond}) should be $l\pi, l\in\Z$. The computation of root-locus trajectories for boundary crossing roots remains the same. The poles of $G$ inside $\Cs$ are traced in two steps, first from $k=0$ to $k=k_{\max}$, then from $k=0$ to $k=-k_{\max}$.

\noindent\textbf{Remark 2:\ } The algorithm can be extended to the case where $\Re(s)=\so>0$. The value of $k_{\max}$ can be chosen such that the asymptotic properties of the roots are observed.
\section{Example} \label{sec:ex}
\begin{figure}[b]
\begin{center}
\includegraphics[width=7cm]{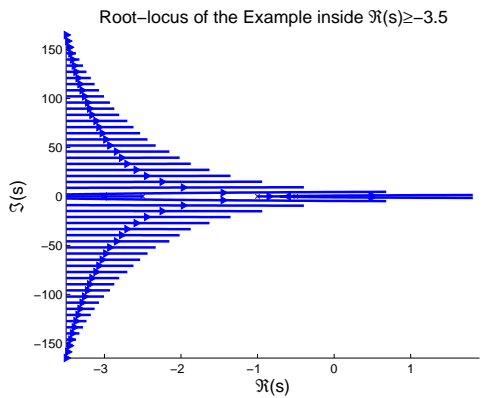}
\caption{\label{fig:rlocus} The root-locus trajectories inside $\Re(s)\geq-3.5$ }
\end{center}
\end{figure}

\begin{figure}[t]
\begin{center}
\includegraphics[width=7cm]{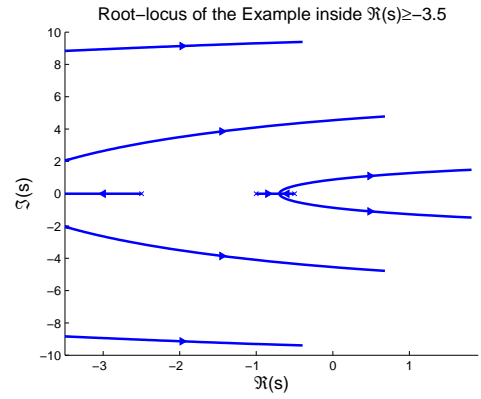}
\caption{\label{fig:rlocus_zoomed} The root-locus trajectories (zoomed)}
\end{center}
\end{figure}

We consider the following SISO dead-time system (\ref{eq:tfG})
\bd
{\textstyle \left(\frac{s^2-10s+50}{s^3 + 4 s^2 + 4.25 s + 1.25}\right)e^{-s}.}
\ed
The nominal contraction rate and distance are set to $\tilde{\kappa}=1.1$, $\tilde{\delta}=10^{-3}$ and the tolerance for the corrector step is $10^{-6}$. The root-locus trajectories inside the root-locus region $\Re(s)\geq-3.5$ for the controller gain interval $k\in[0,5]$ are given in Figure~\ref{fig:rlocus}. The boundary crossing roots enter the root-locus region and their corresponding trajectories can be seen. The root-locus trajectories of the poles of $G$ inside the root-locus region, $s=-0.5, -1, -2.5$, are shown in Figure~\ref{fig:rlocus_zoomed}. The trajectories of $s=-0.5$ and $s=-0.1$ have a branching point at $s=-0.7$, then they converge to the zeros of $G$, $s=5\pm5i$. The trajectory of $s=-2.5$ leaves the root-locus region.

\vspace{-.2cm}
\section{Concluding Remarks} \label{sec:concl}\vspace{-.3cm}
A continuation method to compute the root-locus of SISO dead-time systems within a root-locus region, a given right complex half-plane, is given. The method calculates the starting points of root-locus trajectories including the ones crossing the boundary of root-locus region. The roots on each root-locus trajectory are predicted by a secant method where the step length is adaptive and the predicted values are corrected by Newton's method. The implementation is numerically stable and effective for high-order systems.
\vspace{-.3cm}
\bibliography{rlocus}
\end{document}